\documentclass[aip,apl,10pt,twocolumn]{revtex4-1}

\usepackage{graphicx}
\usepackage{amsmath}
\usepackage{xspace}

\newcommand{\aop}{\ensuremath{a^{\phantom{\dagger}}}}
\newcommand{\bop}{\ensuremath{b^{\phantom{\dagger}}}}
\newcommand{\cop}{\ensuremath{c^{\phantom{\dagger}}}}

\newcommand{\HSP}{\ensuremath{H_{\mathrm{SP}}}}
\newcommand{\Vinter}{\ensuremath{V_{\mathrm{e-h}}}\xspace}

\newcommand{\Eext}{\ensuremath{E^{\mathrm{ext}}}\xspace}
\newcommand{\mast}{\ensuremath{m^{\ast}}\xspace}

\newcommand{\Dmax}{\ensuremath{\Delta_{\mathrm{max}}}\xspace}
\newcommand{\mucrit}{\ensuremath{\mu_{\mathrm{crit}}}\xspace}
\newcommand{\eV}{\ensuremath{\mathrm{eV}}\xspace}

\begin{document}

\title{Excitonic condensation in spatially separated one-dimensional
systems}
\author{D.\,S.\,L.~Abergel}
\affiliation{Nordita, KTH Royal Institute of Technology and Stockholm
University, Roslagstullsbacken 23, SE-106 91 Stockholm, Sweden.}
\affiliation{Center for Quantum Materials, KTH and Nordita,
Roslagstullsbacken 17, SE-106 91 Stockholm, Sweden.}

\begin{abstract}
We show theoretically that excitons can form from spatially separated
one-dimensional ground state populations of electrons and holes, and
that the resulting excitons can form a quasicondensate.
We describe a mean-field Bardeen-Cooper-Schrieffer theory in the low
carrier density regime and then focus on the core-shell nanowire giving
estimates of the size of the excitonic gap for InAs/GaSb wires and as a
function of all the experimentally relevant parameters.
We find that optimal conditions for pairing include small overlap of the
electron and hole bands, large effective mass of the carriers, and low
dielectric constant of the surrounding media.
Therefore, one-dimensional systems provide an attractive platform for
the experimental detection of excitonic quasicondensation in zero
magnetic field.
\end{abstract}

\maketitle

Excitonic condensation is the formation of a macroscopic condensed state
of bosons consisting of paired electrons and holes.
\cite{Lozovik-JETP44}
The quasiparticles which comprise this boson may come either from
optically pumped (and hence transient) states, or from spatially
separated ground state populations which are stable for, in principle,
infinite time.
In addition to being a highly interesting physical phenomenon, the
excitonic condensate may have important technological applications such
as in dispersionless switching devices,\cite{Banerjee-EDL30} 
an analogue-to-digital converter,\cite{Dolcini-PRL104}
and for the design of topologically protected qubits.
\cite{Peotta-PRB84}
Spatially separated exciton condensates have a long history of
consideration in two dimensions (2D), where electron-hole bilayers made
first from semiconductor heterostructures,\cite{Lozovik-JETP44} and
later from two graphene layers \cite{Zhang-PRB77,Min-PRB78} were
proposed as systems where this phenomenon could be observed. 
However, the only clear experimental evidence for such a state has been
seen in a strong magnetic field,\cite{Eisenstein-Nature432,
Nandi-Nature488} and despite much effort, the excitonic condensate has
not been observed in zero field.
\cite{Pillarisetty-PRL89, Seamons-PRL102, Kim-PRB83, Gorbachev-NatPhys8,
Gamucci-NatComms5} 
Theoretical estimates indicate that monolayer graphene systems with
conical band structures are probably too weakly interacting for the
condensate to manifest at zero field,
\cite{Sodemann-PRB85, Lozovik-PRB86} but that multilayer graphene
systems may be an attractive platform.\cite{Zhang-PRL111,
Perali-PRL110, Zarenia-SciRep4}

\begin{figure}[tb]
	\centering
	\includegraphics[]{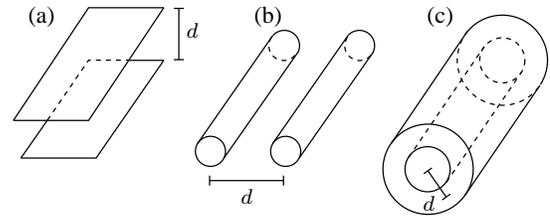}
	\caption{Three candidate experimental systems. (a) Double layer
	geometry such as stacked graphene nanoribbons or a InAs/GaSb double
	quantum well, (b) side-by-side nanowires or carbon nanotubes, 
	and (c) a core-shell nanowire. The effective layer separation $d$ is
	marked in each case.
	\label{fig:systems}}
\end{figure}

%% Experimental systems
Such excitonic condensation formally cannot exist in one dimension (1D)
in the true sense of having long-range order, but correlations between
excitons may exist at finite length scales and are characterised by a
power law decay. \cite{Cazalilla-RMP83, Guan-RMP85}
We propose that this `quasicondensation' may be supported in various
experimentally relevant condensed matter systems 
\footnote{Additionally, constructing quasi-1D systems of superconducting
metals has been shown to enhance the critical temperature with respect
to the bulk system. See A.~Bianconi, A.~Valletta, A.~Perali, and
N.~L.~Saini, Physica C: Supercond.~\textbf{296}, 269 (1998),
A.~A.~Shanenko, M.~D.~Croitoru, M.~Zgirski, F.~M.~Peeters, and
K.~Arutyunov, Phys.~Rev.~B~\textbf{74}, 052502 (2006) and references
therein.}
including stacked graphene nanoribbons, \cite{Brey-PRB73} in nanowires
or carbon nanotubes arranged side-by-side,\cite{arXiv1408.2718} in a
core-shell nanowire (CSN), \cite{Ganjipour-APL101} in nanowires
constructed from adjacent electron-doped and hole-doped GaAs quantum
wells, or in nanowires patterned from a InAs/GaSb double quantum well. 
These setups are illustrated schematically in Fig.~\ref{fig:systems}.
In all candidate systems, the important experimental 
prerequisites are that the electron and hole populations are
simultaneously present in the ground state and separated by a finite
distance $d$.
If transport measurements are employed to detect the condensate, then 
independent contacting of the two layers is essential to allow
either Coulomb drag measurements,\cite{Su-NatPhys4} or measurement of
the tunneling current between the two layers.\cite{Eisenstein-Nature432}
In the case of stacked graphene, this has been demonstrated many times
in 2D since the two layers can be gated from opposite sides
\cite{Gorbachev-NatPhys8} and contacting the two layers is also
straightforward.
The growth process of CSNs will also allow this, since it is possible to
grow the shell around only some length of the core and hence leave both
materials exposed for independent contacting.
Alternatively, optical probes could be used which do not require
contacts.

In this Letter, we present the mean-field BCS theory that is applicable
for all spatially separated 1D electron and hole systems in the limit of
low carrier density. 
We then give numerical results which illustrate the behavior of a CSN as
it depends on the material parameters.
Finally, we discuss some specific practical details of CSN systems, and
how excitonic quasicondensation would differ in the other possible
setups.  
Previously, excitonic condensation has been considered in 1D in the
context of the high density limit of optical excitations,
\cite{Nagaosa-SSC88} BCS theory was applied to cold fermionic gases
generally using a short-range interaction,\cite{Schlottmann-JPCM6,
Bruun-EPJD7, Bedaque-PRL91, Tokatly-PRL93, Ying-PRL100}
in carbon nanotubes,\cite{Bondarev-PRB89}
and in a nonequilibrium context.\cite{Zachmann-NJP15}
However, spatially separated solid state systems have only been
investigated for excitonic quasicondensation once,\cite{arXiv1408.2718}
where the language of bosonization was used and only qualitative
predictions for the interwire tunneling current were given.

%% Single particle theory
The general theory consists of a Hamiltonian $H = \HSP + V$ where
$\HSP$ contains the single particle terms and $V$ is the
interwire interaction. 
We assume that both the electron and hole regions contain one
nondegenerate band and allow for independent chemical potentials in
each region.
\footnote{For a realistic spin-degenerate material this one band
approximation can be realised using the Zeeman splitting associated with
a magnetic field (since Landau levels will not occur in 1D) or with a
ferromagnetic substrate.}
The annililation operator for an electron in one of these bands is
denoted $\cop_{bk}$ where $b=1,2$ indicates the band, and $k$ is
the momentum. Since we enforce one band to be electron-like and the
other to be hole-like we introduce the following notation: $\aop_k =
\cop_{1k}$ is the annihilation of an electron in layer 1, 
and $\bop_{-k} = c^\dagger_{2k}$ is the annihilation of a hole in layer 2.
The single particle part of the Hamiltonian is
\begin{equation*}
	\HSP = 
	\sum_k \xi_{1k}  a^\dagger_{k} \aop_{k}
	+ \sum_k \xi_{2k} \bop_{-k} b^\dagger_{-k}.
\end{equation*}
with $\xi_{1k} = \epsilon_{1k} - \mu_1$ and $\xi_{2k} = \epsilon_{2k} -
\mu_2$.
This Hamiltonian does not include any interwire tunneling but we
qualitatively discuss its effect below. Also, in principle a realistic
1D system will contain many subbands, but in the low density limit we
assume that only the first subband plays an active role in the excitonic
pairing.

The interwire interaction between the electrons and holes is derived
within mean-field BCS theory. \cite{Liu-PRA76, Parish-PRL99,
Edge-PRL103}
This is valid at low temperature (when $T$ is much less than the
mean-field critical temperature $T_c$) and when the interlayer
interaction is sufficiently strong, \cite{arXiv1408.2718} corresponding
to the limit where separation of the two carrier species $d$ is less
than the average separation of carriers within each wire given by the
inverse of the carrier density, so that the instability towards exciton
formation dominates over the tendency to form a Luttinger liquid state
within each wire. 
For this to be the case, we require $k_F d < \pi$ (where $k_F$ is the
Fermi wave vector), and for physically realistic parameters and a
quadratic band, this condition is satisfied for $\xi_1 \sim \xi_2 \sim
100\mathrm{meV}$.
The resulting interaction Hamiltonian is $V = \sum_k \Delta_k
a^\dagger_k b^\dagger_{-k} + \mathrm{h.c}$ where $\Delta_k$ is defined
as the self-consistent solution of the equation
\begin{equation}
	\Delta_k = \int dk' \frac{\Vinter(k'-k)}{4\pi}
	\frac{\Delta_{k'}\left[ n_a(k') + n_b(k')-1 \right]}%
	{\sqrt{(\xi_1 - \xi_2)^2 + 4\Delta_{k'}^2}},
	\label{eq:Delta}
\end{equation}
where $n_{a,b}(k)$ are the occupation factors for the bands at wave
vector $k$.
%% Interaction
The quantity \Vinter is the Fourier transform of the interwire
interaction and is given by
\begin{equation*}
	\Vinter(q) = -\frac{2 e^2}{\kappa} K_0(|q|d) 
\end{equation*}
where $K_0(z)$ is the modified Bessel function of the second kind
\footnote{This Bessel function diverges as $z\to 0$, so we impose a
	minimum cutoff $|q|d = 10^{-6}$. We have tested that the precise
value of this cutoff does not impact the numerical results.} and
$\kappa = 4\pi\epsilon_0 \epsilon_r$ in SI units.
The minus sign appears because each 1D system contains carriers of
opposite charge so that their Coulomb interaction is attractive.
In principle, a form factor can also appear in
Eq.~\eqref{eq:Delta}, but this typically produces only a small
quantitative change in the gap function. Also, if the mismatch in
dielectric constants between the core and shell region is large, this may
contribute a substantial renormalisation of the excition binding energy.
\cite{Slachmuylders-PRB74}
This Hamiltonian is diagonalized using a Bogolyuobov transformation
which yields two excitonic bands with dispersion
\begin{equation*}
	E_{\pm,k} = \pm\frac{\xi_{1k} + \xi_{2k}}{2} +
	\frac{1}{2} \sqrt{ (\xi_{1k}-\xi_{2k})^2 + 4\Delta_k^2 }.
\end{equation*}
This indicates that, for finite $\Delta_k$,  a gap opens in the single
particle spectrum near the points where the two single particle bands
cross, and the magnitude of this gap is exactly $\Delta_k$ evaluated at
the wave vector of the crossing point.
Therefore, this value represents the turning point of each of the
excitonic bands $E_{\pm,k}$ and hence the overall gap of these bands
from the excitonic level. Hence, it is called the `excitonic gap', and
label it by $\Dmax$.

However, to give a concrete example and experimentally relevant
predictions, we make our theory specific to the CSN. 
The CSN is a particularly interesting case since it has recently been
demonstrated that certain combinations of core and shell materials allow
for simultaneous populations of holes in the core region and electrons in
the shell region, minimal hybridization between them, and ambipolar
transport characteristics. \cite{Ganjipour-APL101}
Here, it is reasonable to approximate the low energy band structure in a
quadratic form \cite{Pistol-PRB78,Kishore-PRB86}
parameterized by a band offset $\Eext$ and an effective mass $\mast$
such that
\begin{equation*}
	\epsilon_{ik} = \Eext_i + \frac{\hbar^2 k^2}{2\mast_i}.
\end{equation*}
One important combination of materials is a CSN with a core (hole)
region of GaSb and a shell (electron) region of InAs.
Using parameters from Ref.~\onlinecite{Pistol-PRB78}, we take band
extrema of $\Eext_{\mathrm{InAs}}=-0.2\eV$ and
$\Eext_{\mathrm{GaSb}}=-0.05\eV$, and effective masses of
$\mast_{\mathrm{InAs}}=0.02$ and $\mast_{\mathrm{GaSb}}=-0.07$.
This pair of materials fulfills the general requirements for the
excitons to exist: they have bands that overlap with opposite sign of the
effective mass.
We assume a dielectric constant of $\epsilon_r = 12$ in both materials
so that the dielectric mismatch is zero and there is no renormalisation
of the exciton binding energy.

\begin{figure}
	\centering
	\includegraphics[]{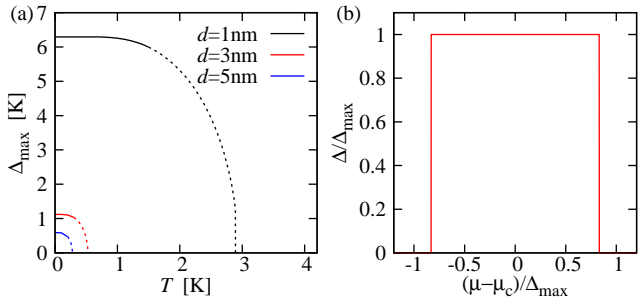}
	\caption{(a) Predicted \Dmax as a function of temperature for
	an InAs/GaSb CSN for several different layer separations, and
	$\mu=\mu_c$.
	(b) Dependence of \Dmax on the chemical potential $\mu$ for a
	InAs/GaSb CSN at $T=0$.
	\label{fig:InAsGaSb}}
\end{figure}

The pairing happens most strongly when the chemical potential
$\mu=\mu_1=\mu_2$ is placed at the point where the two bands cross,
given by $\mu_c = (\mast_1 \Eext_1 - \mast_2 \Eext_2)/(\mast_1 -
\mast_2)$.
Figure \ref{fig:InAsGaSb}(a) shows the predicted maximum of the gap
function \Dmax for a InAs/GaSb CSN as function of temperature for
various inter-wire spacings and $\mu=\mu_c$. We stress that the
mean-field BCS will not give a quantitatively accurate estimate of the
critical temperature ($T_c$) so the higher-temperature part is shown as
a dotted line to emphasize this.
The estimated gap size is of the order of one milli-electron volt at
$T=0$, so that the
associated $T_c$ is of the order of a few Kelvin, and is strongly
dependent on the spacing between the electron and hole populations. 
Figure \ref{fig:InAsGaSb}(b) shows how $\Dmax$ behaves
as a function of the chemical potential at $T=0$. 
As $\mu$ moves away from the
band crossing at $\mu_c$, the gap stays unchanged until a certain point
where it collapses to zero immediately. 
This is caused by the minimum of one of the excitonic bands $E_{\pm,k}$
becoming negative and allowing for gapless excitations at the Fermi
energy which are energetically favorable to the exciton state, in close
analogy to the Clogston-Chandrasekhar limit for a superconductor in a
magnetic field.
The value of the chemical potential for which this happens at $T=0$,
(which we label $\mucrit$), is 
\begin{equation}
	\mucrit = \mu_c \pm 2\Dmax
	\frac{\sqrt{|\mast_1||\mast_2|}}{|\mast_1-\mast_2|}.
	\label{eq:mux}
\end{equation}
For the parameters appropriate to the InAs/GaSb CSN we find $\mucrit =
\mu_c \pm 0.831\Dmax$. 
The coefficient of \Dmax in the second term is peaked when $\mast_1 =
-\mast_2$ indicating that the exciton is the most robust against
shifts in the chemical potential when the band masses have equal
magnitude.

\begin{figure}[tb]
	\centering
	\includegraphics[]{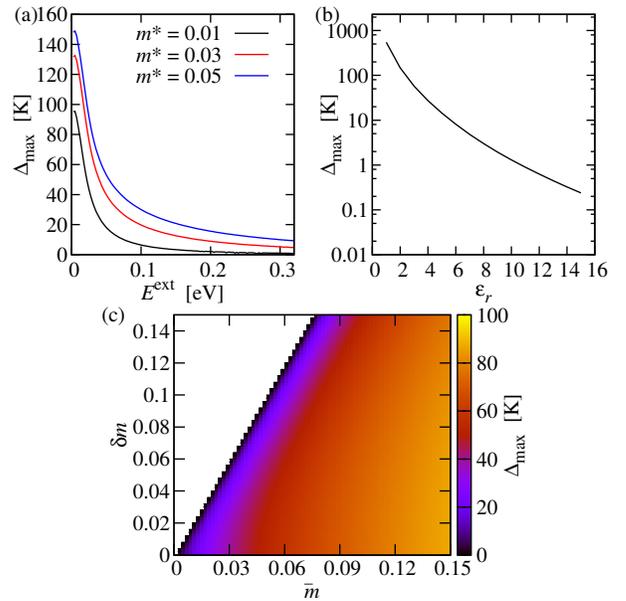}
	\caption{Effect of band parameters on the size of the excitonic gap
	in a CSN.
	(a) \Dmax as a function of the band overlap for different effective
	masses.
	We fix $d=5\mathrm{nm}$ and $\epsilon_r=4$.
	(b) \Dmax as a function of $\epsilon_r$.  We use $d=5\mathrm{nm}$
	and band parameters suitable for an InAs/GaSb wire.
	(c) \Dmax as a function of $\bar{m}$ and $\delta m$ with
	$d=5\mathrm{nm}$, $\epsilon_r = 4$, and $\Eext=0.05\eV$.
	\label{fig:bandpars}}
\end{figure}

In principle, the choice of material for the core and shell regions
could be made so as to optimize the excitonic gap.
Therefore, we analyze \Dmax at $T=0$ as a function of the band
parameters and dielectric constant.
Figure \ref{fig:bandpars}(a) shows \Dmax as a function of the overlap of
the two bands such that $\Eext = -\Eext_1 = \Eext_2$ for three different
values of the effective masses defined such that $\mast = \mast_1 =
-\mast_2$.
We fix all other parameters. 
This indicates that the pairing is strongest when $\mu_c$ is close to
the band extrema so that the Fermi surface is small and the carrier
density is low.
Figure \ref{fig:bandpars}(c) shows \Dmax as a function of the band
masses.
Here, we arrange the notation such that $\bar{m} =
(|\mast_1|+|\mast_2|)/2$ and $\delta m = |\mast_1| - |\mast_2|$ while
fixing the criteria that $\mast_1>0$ and $\mast_2<0$.
The unfilled region in the upper-left part of the plot is the parameter
space where these criteria cannot be satisfied. 
Looking first along the $\delta m = 0$ axis, we see that \Dmax increases
with the band masses. 
This, taken together with the observations from
Fig.~\ref{fig:bandpars}(a) indicate that flatter bands (and hence a higher
density of states) increases the excitonic gap. 
Allowing $\delta m$ to be finite shows that imbalance in the two
effective masses is detrimental to the excitonic pairing, although this
effect diminishes as $\bar{m}$ increases.
Figure \ref{fig:bandpars}(b) shows that the dielectric constant of the
medium is also a crucial parameter. 
We find that suppression of the excitonic gap with increasing
$\epsilon_r$ is slightly sub-exponential. 

%% Band parameters

%% Hybridization
Since we focus on the CSN system where there is no
spacer layer between the electron and hole regions, we have to take the
possible direct tunneling of carriers between the two bands, and the
resulting hybridization of the single particle bands into account. The
hybridization of bands in the context of 2D excitonic condensation has
been analyzed previously \cite{Sodemann-PRB85,Efimkin-PRB86} where it
was found that a small hybridization was beneficial to the formation of
a gap at the Fermi energy, since it adds a single particle contribution
to the $a^\dagger_k b^\dagger_{-k}$ channel. 
In fact, some finite amount of interwire hybridization is essential for
the operation of the BiSFET device \cite{Banerjee-EDL30} and hence the
CSN provides an excellent platform for its implementation.

%% Graphene
The situation with stacked graphene nanoribbons will be qualitatively
similar to that discussed for CSNs. For armchair nanoribbons, the
dispersion associated with the single particle bands will change to
$\epsilon_{ik} = \pm \hbar v_F \sqrt{k_{0i}^2 + k^2}$ where $v_F$ is the
Fermi velocity of the Dirac cone of the graphene, the sign reflects the
conduction or valence band, and $\hbar v_F k_0$ is the extremum of the
lowest subband and is set by the ribbon geometry.\cite{Brey-PRB73}
A form factor resulting from the sublattice structure of the underlying
graphene also has to be added to Eq.~\eqref{eq:Delta} which
will manifest as a small quantitative reduction in the gap size.
The chemical potential of the two layers may be set independently by
gating the stacked ribbons from above and below, but the most
advantageous situation will occur when the Fermi points in the two
layers are at the same wave vector.

%% Disorder
For graphene systems, it is well known that the crystal lattice
structure is of remarkably high quality, and that the dominant form of
disorder is charged impurities trapped in the environment of the flake.
In 2D, the effect of the fluctuations in the local charge density
resulting from these impurities has been shown to be highly detrimental
to the stability of the condensate.\cite{Abergel-PRB86-excon,Abergel-PRB88}
This will be no different for stacked graphene nanoribbons and so a
result analogous to that in Eq.~\eqref{eq:mux} will characterise the
stability of the 1D quasicondensate against local fluctuations in the
chemical potential. 
For III-V semiconductor systems such as the CSNs, growth-related
disorder such as width fluctuations will be important. However, these
will also generate a local shift in the chemical potential, and so as
long as $\mu_{\mathrm{crit}}$ is not exceeded locally, the exciton
formation will be unaffected.
%However, momentum scattering from short range impurities will not be
%detrimental to the stability of the excitons.\cite{Bistritzer-PRL101}

%% Conclusion
In conclusion, we have demonstated that excitonic pairing and the
resulting quasicondensation has a realistic chance of being observed in
1D systems.
In particular, CSNs provide a likely platform for its observation, with
an excitonic gap of the order of Kelvin for thin InAs/GaSb wires. 
This gap size may be substantially increased by choosing materials with
higher effective mass, decreased band overlap, and reduced dielectric
constant.
Throughout, we assume that the interaction between wires is stronger
than the interaction within each wire, so that the Luttinger liquid
effects are neglected.
This is a strong approximation, but current experimental transport data
on CSNs show no evidence of Luttinger liquid effects.
\cite{Ganjipour-APL101}
In particular, metallic InAs nanowires show improved conductance as the
temperature is lowered, and there is no evidence of suppression of
tunneling into core-shell nanowires at low temperature.\cite{Thelander}

%% Acknowledgements
We wish to thank Martin Leijnse, Claes Thelander, Bahram Ganjipour,
Adrian Kantian, and Alexander Balatsky for helpful discussions. This
work was supported by ERC DM-321031 and by Nordita.

\bibliography{jabref-database,mypapers,arXiv,privcomm}

\begin{thebibliography}{45}%
\makeatletter
\providecommand \@ifxundefined [1]{%
 \@ifx{#1\undefined}
}%
\providecommand \@ifnum [1]{%
 \ifnum #1\expandafter \@firstoftwo
 \else \expandafter \@secondoftwo
 \fi
}%
\providecommand \@ifx [1]{%
 \ifx #1\expandafter \@firstoftwo
 \else \expandafter \@secondoftwo
 \fi
}%
\providecommand \natexlab [1]{#1}%
\providecommand \enquote  [1]{``#1''}%
\providecommand \bibnamefont  [1]{#1}%
\providecommand \bibfnamefont [1]{#1}%
\providecommand \citenamefont [1]{#1}%
\providecommand \href@noop [0]{\@secondoftwo}%
\providecommand \href [0]{\begingroup \@sanitize@url \@href}%
\providecommand \@href[1]{\@@startlink{#1}\@@href}%
\providecommand \@@href[1]{\endgroup#1\@@endlink}%
\providecommand \@sanitize@url [0]{\catcode `\\12\catcode `\$12\catcode
  `\&12\catcode `\#12\catcode `\^12\catcode `\_12\catcode `\%12\relax}%
\providecommand \@@startlink[1]{}%
\providecommand \@@endlink[0]{}%
\providecommand \url  [0]{\begingroup\@sanitize@url \@url }%
\providecommand \@url [1]{\endgroup\@href {#1}{\urlprefix }}%
\providecommand \urlprefix  [0]{URL }%
\providecommand \Eprint [0]{\href }%
\providecommand \doibase [0]{http://dx.doi.org/}%
\providecommand \selectlanguage [0]{\@gobble}%
\providecommand \bibinfo  [0]{\@secondoftwo}%
\providecommand \bibfield  [0]{\@secondoftwo}%
\providecommand \translation [1]{[#1]}%
\providecommand \BibitemOpen [0]{}%
\providecommand \bibitemStop [0]{}%
\providecommand \bibitemNoStop [0]{.\EOS\space}%
\providecommand \EOS [0]{\spacefactor3000\relax}%
\providecommand \BibitemShut  [1]{\csname bibitem#1\endcsname}%
\let\auto@bib@innerbib\@empty
%</preamble>
\bibitem [{\citenamefont {Lozovik}\ and\ \citenamefont
  {Yudson}(1976)}]{Lozovik-JETP44}%
  \BibitemOpen
  \bibfield  {author} {\bibinfo {author} {\bibfnamefont {Y.~E.}\ \bibnamefont
  {Lozovik}}\ and\ \bibinfo {author} {\bibfnamefont {V.}~\bibnamefont
  {Yudson}},\ }\href@noop {} {\bibfield  {journal} {\bibinfo  {journal} {JETP}\
  }\textbf {\bibinfo {volume} {44}},\ \bibinfo {pages} {389} (\bibinfo {year}
  {1976})}\BibitemShut {NoStop}%
\bibitem [{\citenamefont {Banerjee}\ \emph {et~al.}(2009)\citenamefont
  {Banerjee}, \citenamefont {Register}, \citenamefont {Tutuc}, \citenamefont
  {Reddy},\ and\ \citenamefont {MacDonald}}]{Banerjee-EDL30}%
  \BibitemOpen
  \bibfield  {author} {\bibinfo {author} {\bibfnamefont {S.}~\bibnamefont
  {Banerjee}}, \bibinfo {author} {\bibfnamefont {L.}~\bibnamefont {Register}},
  \bibinfo {author} {\bibfnamefont {E.}~\bibnamefont {Tutuc}}, \bibinfo
  {author} {\bibfnamefont {D.}~\bibnamefont {Reddy}}, \ and\ \bibinfo {author}
  {\bibfnamefont {A.}~\bibnamefont {MacDonald}},\ }\href {\doibase
  10.1109/led.2008.2009362} {\bibfield  {journal} {\bibinfo  {journal} {IEEE
  Electron Device Lett.}\ }\textbf {\bibinfo {volume} {30}},\ \bibinfo {pages}
  {158} (\bibinfo {year} {2009})}\BibitemShut {NoStop}%
\bibitem [{\citenamefont {Dolcini}\ \emph {et~al.}(2010)\citenamefont
  {Dolcini}, \citenamefont {Rainis}, \citenamefont {Taddei}, \citenamefont
  {Polini}, \citenamefont {Fazio},\ and\ \citenamefont
  {MacDonald}}]{Dolcini-PRL104}%
  \BibitemOpen
  \bibfield  {author} {\bibinfo {author} {\bibfnamefont {F.}~\bibnamefont
  {Dolcini}}, \bibinfo {author} {\bibfnamefont {D.}~\bibnamefont {Rainis}},
  \bibinfo {author} {\bibfnamefont {F.}~\bibnamefont {Taddei}}, \bibinfo
  {author} {\bibfnamefont {M.}~\bibnamefont {Polini}}, \bibinfo {author}
  {\bibfnamefont {R.}~\bibnamefont {Fazio}}, \ and\ \bibinfo {author}
  {\bibfnamefont {A.~H.}\ \bibnamefont {MacDonald}},\ }\href {\doibase
  10.1103/physrevlett.104.027004} {\bibfield  {journal} {\bibinfo  {journal}
  {Phys. Rev. Lett.}\ }\textbf {\bibinfo {volume} {104}},\ \bibinfo {pages}
  {027004} (\bibinfo {year} {2010})}\BibitemShut {NoStop}%
\bibitem [{\citenamefont {Peotta}\ \emph {et~al.}(2011)\citenamefont {Peotta},
  \citenamefont {Gibertini}, \citenamefont {Dolcini}, \citenamefont {Taddei},
  \citenamefont {Polini}, \citenamefont {Ioffe}, \citenamefont {Fazio},\ and\
  \citenamefont {MacDonald}}]{Peotta-PRB84}%
  \BibitemOpen
  \bibfield  {author} {\bibinfo {author} {\bibfnamefont {S.}~\bibnamefont
  {Peotta}}, \bibinfo {author} {\bibfnamefont {M.}~\bibnamefont {Gibertini}},
  \bibinfo {author} {\bibfnamefont {F.}~\bibnamefont {Dolcini}}, \bibinfo
  {author} {\bibfnamefont {F.}~\bibnamefont {Taddei}}, \bibinfo {author}
  {\bibfnamefont {M.}~\bibnamefont {Polini}}, \bibinfo {author} {\bibfnamefont
  {L.~B.}\ \bibnamefont {Ioffe}}, \bibinfo {author} {\bibfnamefont
  {R.}~\bibnamefont {Fazio}}, \ and\ \bibinfo {author} {\bibfnamefont {A.~H.}\
  \bibnamefont {MacDonald}},\ }\href {\doibase 10.1103/physrevb.84.184528}
  {\bibfield  {journal} {\bibinfo  {journal} {Phys. Rev. B}\ }\textbf {\bibinfo
  {volume} {84}},\ \bibinfo {pages} {184528} (\bibinfo {year}
  {2011})}\BibitemShut {NoStop}%
\bibitem [{\citenamefont {Zhang}\ and\ \citenamefont
  {Joglekar}(2008)}]{Zhang-PRB77}%
  \BibitemOpen
  \bibfield  {author} {\bibinfo {author} {\bibfnamefont {C.-H.}\ \bibnamefont
  {Zhang}}\ and\ \bibinfo {author} {\bibfnamefont {Y.}~\bibnamefont
  {Joglekar}},\ }\href {\doibase 10.1103/physrevb.77.233405} {\bibfield
  {journal} {\bibinfo  {journal} {Phys. Rev. B}\ }\textbf {\bibinfo {volume}
  {77}},\ \bibinfo {pages} {233405} (\bibinfo {year} {2008})}\BibitemShut
  {NoStop}%
\bibitem [{\citenamefont {Min}\ \emph {et~al.}(2008)\citenamefont {Min},
  \citenamefont {Bistritzer}, \citenamefont {Su},\ and\ \citenamefont
  {MacDonald}}]{Min-PRB78}%
  \BibitemOpen
  \bibfield  {author} {\bibinfo {author} {\bibfnamefont {H.}~\bibnamefont
  {Min}}, \bibinfo {author} {\bibfnamefont {R.}~\bibnamefont {Bistritzer}},
  \bibinfo {author} {\bibfnamefont {J.-J.}\ \bibnamefont {Su}}, \ and\ \bibinfo
  {author} {\bibfnamefont {A.}~\bibnamefont {MacDonald}},\ }\href {\doibase
  10.1103/physrevb.78.121401} {\bibfield  {journal} {\bibinfo  {journal} {Phys.
  Rev. B}\ }\textbf {\bibinfo {volume} {78}},\ \bibinfo {pages} {121401}
  (\bibinfo {year} {2008})}\BibitemShut {NoStop}%
\bibitem [{\citenamefont {Eisenstein}\ and\ \citenamefont
  {MacDonald}(2004)}]{Eisenstein-Nature432}%
  \BibitemOpen
  \bibfield  {author} {\bibinfo {author} {\bibfnamefont {J.~P.}\ \bibnamefont
  {Eisenstein}}\ and\ \bibinfo {author} {\bibfnamefont {A.~H.}\ \bibnamefont
  {MacDonald}},\ }\href {\doibase 10.1038/nature03081} {\bibfield  {journal}
  {\bibinfo  {journal} {Nature}\ }\textbf {\bibinfo {volume} {432}},\ \bibinfo
  {pages} {691} (\bibinfo {year} {2004})}\BibitemShut {NoStop}%
\bibitem [{\citenamefont {Nandi}\ \emph {et~al.}(2012)\citenamefont {Nandi},
  \citenamefont {Finck}, \citenamefont {Eisenstein}, \citenamefont {Pfeiffer},\
  and\ \citenamefont {West}}]{Nandi-Nature488}%
  \BibitemOpen
  \bibfield  {author} {\bibinfo {author} {\bibfnamefont {D.}~\bibnamefont
  {Nandi}}, \bibinfo {author} {\bibfnamefont {A.~D.~K.}\ \bibnamefont {Finck}},
  \bibinfo {author} {\bibfnamefont {J.~P.}\ \bibnamefont {Eisenstein}},
  \bibinfo {author} {\bibfnamefont {L.~N.}\ \bibnamefont {Pfeiffer}}, \ and\
  \bibinfo {author} {\bibfnamefont {K.~W.}\ \bibnamefont {West}},\ }\href
  {\doibase 10.1038/nature11302} {\bibfield  {journal} {\bibinfo  {journal}
  {Nature}\ }\textbf {\bibinfo {volume} {488}},\ \bibinfo {pages} {481}
  (\bibinfo {year} {2012})}\BibitemShut {NoStop}%
\bibitem [{\citenamefont {Pillarisetty}\ \emph {et~al.}(2002)\citenamefont
  {Pillarisetty}, \citenamefont {Noh}, \citenamefont {Tsui}, \citenamefont
  {De~Poortere}, \citenamefont {Tutuc},\ and\ \citenamefont
  {Shayegan}}]{Pillarisetty-PRL89}%
  \BibitemOpen
  \bibfield  {author} {\bibinfo {author} {\bibfnamefont {R.}~\bibnamefont
  {Pillarisetty}}, \bibinfo {author} {\bibfnamefont {H.}~\bibnamefont {Noh}},
  \bibinfo {author} {\bibfnamefont {D.}~\bibnamefont {Tsui}}, \bibinfo {author}
  {\bibfnamefont {E.}~\bibnamefont {De~Poortere}}, \bibinfo {author}
  {\bibfnamefont {E.}~\bibnamefont {Tutuc}}, \ and\ \bibinfo {author}
  {\bibfnamefont {M.}~\bibnamefont {Shayegan}},\ }\href {\doibase
  10.1103/physrevlett.89.016805} {\bibfield  {journal} {\bibinfo  {journal}
  {Phys. Rev. Lett.}\ }\textbf {\bibinfo {volume} {89}},\ \bibinfo {pages}
  {016805} (\bibinfo {year} {2002})}\BibitemShut {NoStop}%
\bibitem [{\citenamefont {Seamons}\ \emph {et~al.}(2009)\citenamefont
  {Seamons}, \citenamefont {Morath}, \citenamefont {Reno},\ and\ \citenamefont
  {Lilly}}]{Seamons-PRL102}%
  \BibitemOpen
  \bibfield  {author} {\bibinfo {author} {\bibfnamefont {J.~A.}\ \bibnamefont
  {Seamons}}, \bibinfo {author} {\bibfnamefont {C.~P.}\ \bibnamefont {Morath}},
  \bibinfo {author} {\bibfnamefont {J.~L.}\ \bibnamefont {Reno}}, \ and\
  \bibinfo {author} {\bibfnamefont {M.~P.}\ \bibnamefont {Lilly}},\ }\href
  {\doibase 10.1103/physrevlett.102.026804} {\bibfield  {journal} {\bibinfo
  {journal} {Phys. Rev. Lett.}\ }\textbf {\bibinfo {volume} {102}},\ \bibinfo
  {pages} {026804} (\bibinfo {year} {2009})}\BibitemShut {NoStop}%
\bibitem [{\citenamefont {Kim}\ \emph {et~al.}(2011)\citenamefont {Kim},
  \citenamefont {Jo}, \citenamefont {Nah}, \citenamefont {Yao}, \citenamefont
  {Banerjee},\ and\ \citenamefont {Tutuc}}]{Kim-PRB83}%
  \BibitemOpen
  \bibfield  {author} {\bibinfo {author} {\bibfnamefont {S.}~\bibnamefont
  {Kim}}, \bibinfo {author} {\bibfnamefont {I.}~\bibnamefont {Jo}}, \bibinfo
  {author} {\bibfnamefont {J.}~\bibnamefont {Nah}}, \bibinfo {author}
  {\bibfnamefont {Z.}~\bibnamefont {Yao}}, \bibinfo {author} {\bibfnamefont
  {S.~K.}\ \bibnamefont {Banerjee}}, \ and\ \bibinfo {author} {\bibfnamefont
  {E.}~\bibnamefont {Tutuc}},\ }\href {\doibase 10.1103/physrevb.83.161401}
  {\bibfield  {journal} {\bibinfo  {journal} {Phys. Rev. B}\ }\textbf {\bibinfo
  {volume} {83}},\ \bibinfo {pages} {161401} (\bibinfo {year}
  {2011})}\BibitemShut {NoStop}%
\bibitem [{\citenamefont {Gorbachev}\ \emph {et~al.}(2012)\citenamefont
  {Gorbachev}, \citenamefont {Geim}, \citenamefont {Katsnelson}, \citenamefont
  {Novoselov}, \citenamefont {Tudorovskiy}, \citenamefont {Grigorieva},
  \citenamefont {MacDonald}, \citenamefont {Morozov}, \citenamefont {Watanabe},
  \citenamefont {Taniguchi},\ and\ \citenamefont
  {Ponomarenko}}]{Gorbachev-NatPhys8}%
  \BibitemOpen
  \bibfield  {author} {\bibinfo {author} {\bibfnamefont {R.~V.}\ \bibnamefont
  {Gorbachev}}, \bibinfo {author} {\bibfnamefont {A.~K.}\ \bibnamefont {Geim}},
  \bibinfo {author} {\bibfnamefont {M.~I.}\ \bibnamefont {Katsnelson}},
  \bibinfo {author} {\bibfnamefont {K.~S.}\ \bibnamefont {Novoselov}}, \bibinfo
  {author} {\bibfnamefont {T.}~\bibnamefont {Tudorovskiy}}, \bibinfo {author}
  {\bibfnamefont {I.~V.}\ \bibnamefont {Grigorieva}}, \bibinfo {author}
  {\bibfnamefont {A.~H.}\ \bibnamefont {MacDonald}}, \bibinfo {author}
  {\bibfnamefont {S.~V.}\ \bibnamefont {Morozov}}, \bibinfo {author}
  {\bibfnamefont {K.}~\bibnamefont {Watanabe}}, \bibinfo {author}
  {\bibfnamefont {T.}~\bibnamefont {Taniguchi}}, \ and\ \bibinfo {author}
  {\bibfnamefont {L.~A.}\ \bibnamefont {Ponomarenko}},\ }\href {\doibase
  10.1038/nphys2441} {\bibfield  {journal} {\bibinfo  {journal} {Nature Phys.}\
  }\textbf {\bibinfo {volume} {8}},\ \bibinfo {pages} {896} (\bibinfo {year}
  {2012})}\BibitemShut {NoStop}%
\bibitem [{\citenamefont {Gamucci}\ \emph {et~al.}(2014)\citenamefont
  {Gamucci}, \citenamefont {Spirito}, \citenamefont {Carrega}, \citenamefont
  {Karmakar}, \citenamefont {Lombardo}, \citenamefont {Bruna}, \citenamefont
  {Pfeiffer}, \citenamefont {West}, \citenamefont {Ferrari}, \citenamefont
  {Polini},\ and\ \citenamefont {et~al.}}]{Gamucci-NatComms5}%
  \BibitemOpen
  \bibfield  {author} {\bibinfo {author} {\bibfnamefont {A.}~\bibnamefont
  {Gamucci}}, \bibinfo {author} {\bibfnamefont {D.}~\bibnamefont {Spirito}},
  \bibinfo {author} {\bibfnamefont {M.}~\bibnamefont {Carrega}}, \bibinfo
  {author} {\bibfnamefont {B.}~\bibnamefont {Karmakar}}, \bibinfo {author}
  {\bibfnamefont {A.}~\bibnamefont {Lombardo}}, \bibinfo {author}
  {\bibfnamefont {M.}~\bibnamefont {Bruna}}, \bibinfo {author} {\bibfnamefont
  {L.~N.}\ \bibnamefont {Pfeiffer}}, \bibinfo {author} {\bibfnamefont {K.~W.}\
  \bibnamefont {West}}, \bibinfo {author} {\bibfnamefont {A.~C.}\ \bibnamefont
  {Ferrari}}, \bibinfo {author} {\bibfnamefont {M.}~\bibnamefont {Polini}}, \
  and\ \bibinfo {author} {\bibnamefont {et~al.}},\ }\href {\doibase
  10.1038/ncomms6824} {\bibfield  {journal} {\bibinfo  {journal} {Nature
  Commun.}\ }\textbf {\bibinfo {volume} {5}},\ \bibinfo {pages} {5824}
  (\bibinfo {year} {2014})}\BibitemShut {NoStop}%
\bibitem [{\citenamefont {Sodemann}, \citenamefont {Pesin},\ and\ \citenamefont
  {MacDonald}(2012)}]{Sodemann-PRB85}%
  \BibitemOpen
  \bibfield  {author} {\bibinfo {author} {\bibfnamefont {I.}~\bibnamefont
  {Sodemann}}, \bibinfo {author} {\bibfnamefont {D.~A.}\ \bibnamefont {Pesin}},
  \ and\ \bibinfo {author} {\bibfnamefont {A.~H.}\ \bibnamefont {MacDonald}},\
  }\href {\doibase 10.1103/PhysRevB.85.195136} {\bibfield  {journal} {\bibinfo
  {journal} {Phys. Rev. B}\ }\textbf {\bibinfo {volume} {85}},\ \bibinfo
  {pages} {195136} (\bibinfo {year} {2012})}\BibitemShut {NoStop}%
\bibitem [{\citenamefont {Lozovik}, \citenamefont {Ogarkov},\ and\
  \citenamefont {Sokolik}(2012)}]{Lozovik-PRB86}%
  \BibitemOpen
  \bibfield  {author} {\bibinfo {author} {\bibfnamefont {Y.~E.}\ \bibnamefont
  {Lozovik}}, \bibinfo {author} {\bibfnamefont {S.~L.}\ \bibnamefont
  {Ogarkov}}, \ and\ \bibinfo {author} {\bibfnamefont {A.~A.}\ \bibnamefont
  {Sokolik}},\ }\href {\doibase 10.1103/PhysRevB.86.045429} {\bibfield
  {journal} {\bibinfo  {journal} {Phys. Rev. B}\ }\textbf {\bibinfo {volume}
  {86}},\ \bibinfo {pages} {045429} (\bibinfo {year} {2012})}\BibitemShut
  {NoStop}%
\bibitem [{\citenamefont {Zhang}\ and\ \citenamefont
  {Rossi}(2013)}]{Zhang-PRL111}%
  \BibitemOpen
  \bibfield  {author} {\bibinfo {author} {\bibfnamefont {J.}~\bibnamefont
  {Zhang}}\ and\ \bibinfo {author} {\bibfnamefont {E.}~\bibnamefont {Rossi}},\
  }\href {\doibase 10.1103/physrevlett.111.086804} {\bibfield  {journal}
  {\bibinfo  {journal} {Phys. Rev. Lett.}\ }\textbf {\bibinfo {volume} {111}},\
  \bibinfo {pages} {086804} (\bibinfo {year} {2013})}\BibitemShut {NoStop}%
\bibitem [{\citenamefont {Perali}, \citenamefont {Neilson},\ and\ \citenamefont
  {Hamilton}(2013)}]{Perali-PRL110}%
  \BibitemOpen
  \bibfield  {author} {\bibinfo {author} {\bibfnamefont {A.}~\bibnamefont
  {Perali}}, \bibinfo {author} {\bibfnamefont {D.}~\bibnamefont {Neilson}}, \
  and\ \bibinfo {author} {\bibfnamefont {A.}~\bibnamefont {Hamilton}},\ }\href
  {\doibase 10.1103/physrevlett.110.146803} {\bibfield  {journal} {\bibinfo
  {journal} {Phys. Rev. Lett.}\ }\textbf {\bibinfo {volume} {110}},\ \bibinfo
  {pages} {146803} (\bibinfo {year} {2013})}\BibitemShut {NoStop}%
\bibitem [{\citenamefont {Zarenia}\ \emph {et~al.}(2014)\citenamefont
  {Zarenia}, \citenamefont {Perali}, \citenamefont {Neilson},\ and\
  \citenamefont {Peeters}}]{Zarenia-SciRep4}%
  \BibitemOpen
  \bibfield  {author} {\bibinfo {author} {\bibfnamefont {M.}~\bibnamefont
  {Zarenia}}, \bibinfo {author} {\bibfnamefont {A.}~\bibnamefont {Perali}},
  \bibinfo {author} {\bibfnamefont {D.}~\bibnamefont {Neilson}}, \ and\
  \bibinfo {author} {\bibfnamefont {F.~M.}\ \bibnamefont {Peeters}},\ }\href
  {\doibase 10.1038/srep07319} {\bibfield  {journal} {\bibinfo  {journal} {Sci.
  Rep.}\ }\textbf {\bibinfo {volume} {4}},\ \bibinfo {pages} {7319} (\bibinfo
  {year} {2014})}\BibitemShut {NoStop}%
\bibitem [{\citenamefont {Cazalilla}\ \emph {et~al.}(2011)\citenamefont
  {Cazalilla}, \citenamefont {Citro}, \citenamefont {Giamarchi}, \citenamefont
  {Orignac},\ and\ \citenamefont {Rigol}}]{Cazalilla-RMP83}%
  \BibitemOpen
  \bibfield  {author} {\bibinfo {author} {\bibfnamefont {M.~A.}\ \bibnamefont
  {Cazalilla}}, \bibinfo {author} {\bibfnamefont {R.}~\bibnamefont {Citro}},
  \bibinfo {author} {\bibfnamefont {T.}~\bibnamefont {Giamarchi}}, \bibinfo
  {author} {\bibfnamefont {E.}~\bibnamefont {Orignac}}, \ and\ \bibinfo
  {author} {\bibfnamefont {M.}~\bibnamefont {Rigol}},\ }\href {\doibase
  10.1103/revmodphys.83.1405} {\bibfield  {journal} {\bibinfo  {journal} {Rev.
  Mod. Phys.}\ }\textbf {\bibinfo {volume} {83}},\ \bibinfo {pages} {1405}
  (\bibinfo {year} {2011})}\BibitemShut {NoStop}%
\bibitem [{\citenamefont {Guan}, \citenamefont {Batchelor},\ and\ \citenamefont
  {Lee}(2013)}]{Guan-RMP85}%
  \BibitemOpen
  \bibfield  {author} {\bibinfo {author} {\bibfnamefont {X.-W.}\ \bibnamefont
  {Guan}}, \bibinfo {author} {\bibfnamefont {M.~T.}\ \bibnamefont {Batchelor}},
  \ and\ \bibinfo {author} {\bibfnamefont {C.}~\bibnamefont {Lee}},\ }\href
  {\doibase 10.1103/revmodphys.85.1633} {\bibfield  {journal} {\bibinfo
  {journal} {Rev. Mod. Phys.}\ }\textbf {\bibinfo {volume} {85}},\ \bibinfo
  {pages} {1633} (\bibinfo {year} {2013})}\BibitemShut {NoStop}%
\bibitem [{Note1()}]{Note1}%
  \BibitemOpen
  \bibinfo {note} {Additionally, constructing quasi-1D systems of
  superconducting metals has been shown to enhance the critical temperature
  with respect to the bulk system. See A.~Bianconi, A.~Valletta, A.~Perali, and
  N.~L.~Saini, Physica C: Supercond.~\protect \textbf {296}, 269 (1998),
  A.~A.~Shanenko, M.~D.~Croitoru, M.~Zgirski, F.~M.~Peeters, and K.~Arutyunov,
  Phys.~Rev.~B~\protect \textbf {74}, 052502 (2006) and references
  therein.}\BibitemShut {Stop}%
\bibitem [{\citenamefont {Brey}\ and\ \citenamefont
  {Fertig}(2006)}]{Brey-PRB73}%
  \BibitemOpen
  \bibfield  {author} {\bibinfo {author} {\bibfnamefont {L.}~\bibnamefont
  {Brey}}\ and\ \bibinfo {author} {\bibfnamefont {H.}~\bibnamefont {Fertig}},\
  }\href {\doibase 10.1103/physrevb.73.235411} {\bibfield  {journal} {\bibinfo
  {journal} {Phys. Rev. B}\ }\textbf {\bibinfo {volume} {73}},\ \bibinfo
  {pages} {235411} (\bibinfo {year} {2006})}\BibitemShut {NoStop}%
\bibitem [{\citenamefont {Werman}\ and\ \citenamefont
  {Berg}(2014)}]{arXiv1408.2718}%
  \BibitemOpen
  \bibfield  {author} {\bibinfo {author} {\bibfnamefont {Y.}~\bibnamefont
  {Werman}}\ and\ \bibinfo {author} {\bibfnamefont {E.}~\bibnamefont {Berg}},\
  }\href@noop {} {\bibfield  {journal} {\bibinfo  {journal} {ArXiv e-prints}\ }
  (\bibinfo {year} {2014})},\ \Eprint {http://arxiv.org/abs/1408.2718}
  {arXiv:1408.2718 [cond-mat.mes-hall]} \BibitemShut {NoStop}%
\bibitem [{\citenamefont {Ganjipour}\ \emph {et~al.}(2012)\citenamefont
  {Ganjipour}, \citenamefont {Ek}, \citenamefont {Mattias~Borg}, \citenamefont
  {Dick}, \citenamefont {Pistol}, \citenamefont {Wernersson},\ and\
  \citenamefont {Thelander}}]{Ganjipour-APL101}%
  \BibitemOpen
  \bibfield  {author} {\bibinfo {author} {\bibfnamefont {B.}~\bibnamefont
  {Ganjipour}}, \bibinfo {author} {\bibfnamefont {M.}~\bibnamefont {Ek}},
  \bibinfo {author} {\bibfnamefont {B.}~\bibnamefont {Mattias~Borg}}, \bibinfo
  {author} {\bibfnamefont {K.~A.}\ \bibnamefont {Dick}}, \bibinfo {author}
  {\bibfnamefont {M.-E.}\ \bibnamefont {Pistol}}, \bibinfo {author}
  {\bibfnamefont {L.-E.}\ \bibnamefont {Wernersson}}, \ and\ \bibinfo {author}
  {\bibfnamefont {C.}~\bibnamefont {Thelander}},\ }\href {\doibase
  10.1063/1.4749283} {\bibfield  {journal} {\bibinfo  {journal} {Appl. Phys.
  Lett.}\ }\textbf {\bibinfo {volume} {101}},\ \bibinfo {pages} {103501}
  (\bibinfo {year} {2012})}\BibitemShut {NoStop}%
\bibitem [{\citenamefont {Su}\ and\ \citenamefont
  {MacDonald}(2008)}]{Su-NatPhys4}%
  \BibitemOpen
  \bibfield  {author} {\bibinfo {author} {\bibfnamefont {J.-J.}\ \bibnamefont
  {Su}}\ and\ \bibinfo {author} {\bibfnamefont {A.~H.}\ \bibnamefont
  {MacDonald}},\ }\href {\doibase 10.1038/nphys1055} {\bibfield  {journal}
  {\bibinfo  {journal} {Nature Phys.}\ }\textbf {\bibinfo {volume} {4}},\
  \bibinfo {pages} {799} (\bibinfo {year} {2008})}\BibitemShut {NoStop}%
\bibitem [{\citenamefont {Nagaosa}\ and\ \citenamefont
  {Ogawa}(1993)}]{Nagaosa-SSC88}%
  \BibitemOpen
  \bibfield  {author} {\bibinfo {author} {\bibfnamefont {N.}~\bibnamefont
  {Nagaosa}}\ and\ \bibinfo {author} {\bibfnamefont {T.}~\bibnamefont
  {Ogawa}},\ }\href {\doibase 10.1016/0038-1098(93)90512-l} {\bibfield
  {journal} {\bibinfo  {journal} {Solid State Commun.}\ }\textbf {\bibinfo
  {volume} {88}},\ \bibinfo {pages} {295} (\bibinfo {year} {1993})}\BibitemShut
  {NoStop}%
\bibitem [{\citenamefont {Schlottmann}(1994)}]{Schlottmann-JPCM6}%
  \BibitemOpen
  \bibfield  {author} {\bibinfo {author} {\bibfnamefont {P.}~\bibnamefont
  {Schlottmann}},\ }\href {\doibase 10.1088/0953-8984/6/20/011} {\bibfield
  {journal} {\bibinfo  {journal} {J. Phys.: Condens. Matter}\ }\textbf
  {\bibinfo {volume} {6}},\ \bibinfo {pages} {3719} (\bibinfo {year}
  {1994})}\BibitemShut {NoStop}%
\bibitem [{\citenamefont {Bruun}\ \emph {et~al.}(1999)\citenamefont {Bruun},
  \citenamefont {Castin}, \citenamefont {Dum},\ and\ \citenamefont
  {Burnett}}]{Bruun-EPJD7}%
  \BibitemOpen
  \bibfield  {author} {\bibinfo {author} {\bibfnamefont {G.}~\bibnamefont
  {Bruun}}, \bibinfo {author} {\bibfnamefont {Y.}~\bibnamefont {Castin}},
  \bibinfo {author} {\bibfnamefont {R.}~\bibnamefont {Dum}}, \ and\ \bibinfo
  {author} {\bibfnamefont {K.}~\bibnamefont {Burnett}},\ }\href@noop {}
  {\bibfield  {journal} {\bibinfo  {journal} {Euro. Phys. J. D}\ }\textbf
  {\bibinfo {volume} {7}},\ \bibinfo {pages} {433} (\bibinfo {year}
  {1999})}\BibitemShut {NoStop}%
\bibitem [{\citenamefont {Bedaque}, \citenamefont {Caldas},\ and\ \citenamefont
  {Rupak}(2003)}]{Bedaque-PRL91}%
  \BibitemOpen
  \bibfield  {author} {\bibinfo {author} {\bibfnamefont {P.}~\bibnamefont
  {Bedaque}}, \bibinfo {author} {\bibfnamefont {H.}~\bibnamefont {Caldas}}, \
  and\ \bibinfo {author} {\bibfnamefont {G.}~\bibnamefont {Rupak}},\ }\href
  {\doibase 10.1103/physrevlett.91.247002} {\bibfield  {journal} {\bibinfo
  {journal} {Phys. Rev. Lett.}\ }\textbf {\bibinfo {volume} {91}},\ \bibinfo
  {pages} {247002} (\bibinfo {year} {2003})}\BibitemShut {NoStop}%
\bibitem [{\citenamefont {Tokatly}(2004)}]{Tokatly-PRL93}%
  \BibitemOpen
  \bibfield  {author} {\bibinfo {author} {\bibfnamefont {I.~V.}\ \bibnamefont
  {Tokatly}},\ }\href {\doibase 10.1103/physrevlett.93.090405} {\bibfield
  {journal} {\bibinfo  {journal} {Phys. Rev. Lett.}\ }\textbf {\bibinfo
  {volume} {93}},\ \bibinfo {pages} {090405} (\bibinfo {year}
  {2004})}\BibitemShut {NoStop}%
\bibitem [{\citenamefont {Ying}\ \emph {et~al.}(2008)\citenamefont {Ying},
  \citenamefont {Cuoco}, \citenamefont {Noce},\ and\ \citenamefont
  {Zhou}}]{Ying-PRL100}%
  \BibitemOpen
  \bibfield  {author} {\bibinfo {author} {\bibfnamefont {Z.-J.}\ \bibnamefont
  {Ying}}, \bibinfo {author} {\bibfnamefont {M.}~\bibnamefont {Cuoco}},
  \bibinfo {author} {\bibfnamefont {C.}~\bibnamefont {Noce}}, \ and\ \bibinfo
  {author} {\bibfnamefont {H.-Q.}\ \bibnamefont {Zhou}},\ }\href {\doibase
  10.1103/physrevlett.100.140406} {\bibfield  {journal} {\bibinfo  {journal}
  {Phys. Rev. Lett.}\ }\textbf {\bibinfo {volume} {100}},\ \bibinfo {pages}
  {140406} (\bibinfo {year} {2008})}\BibitemShut {NoStop}%
\bibitem [{\citenamefont {Bondarev}\ and\ \citenamefont
  {Meliksetyan}(2014)}]{Bondarev-PRB89}%
  \BibitemOpen
  \bibfield  {author} {\bibinfo {author} {\bibfnamefont {I.~V.}\ \bibnamefont
  {Bondarev}}\ and\ \bibinfo {author} {\bibfnamefont {A.~V.}\ \bibnamefont
  {Meliksetyan}},\ }\href {\doibase 10.1103/physrevb.89.045414} {\bibfield
  {journal} {\bibinfo  {journal} {Phys. Rev. B}\ }\textbf {\bibinfo {volume}
  {89}},\ \bibinfo {pages} {045414} (\bibinfo {year} {2014})}\BibitemShut
  {NoStop}%
\bibitem [{\citenamefont {Zachmann}\ \emph {et~al.}(2013)\citenamefont
  {Zachmann}, \citenamefont {Croitoru}, \citenamefont {Vagov}, \citenamefont
  {Axt}, \citenamefont {Papenkort},\ and\ \citenamefont
  {Kuhn}}]{Zachmann-NJP15}%
  \BibitemOpen
  \bibfield  {author} {\bibinfo {author} {\bibfnamefont {M.}~\bibnamefont
  {Zachmann}}, \bibinfo {author} {\bibfnamefont {M.~D.}\ \bibnamefont
  {Croitoru}}, \bibinfo {author} {\bibfnamefont {A.}~\bibnamefont {Vagov}},
  \bibinfo {author} {\bibfnamefont {V.~M.}\ \bibnamefont {Axt}}, \bibinfo
  {author} {\bibfnamefont {T.}~\bibnamefont {Papenkort}}, \ and\ \bibinfo
  {author} {\bibfnamefont {T.}~\bibnamefont {Kuhn}},\ }\href {\doibase
  10.1088/1367-2630/15/5/055016} {\bibfield  {journal} {\bibinfo  {journal}
  {New J. Phys.}\ }\textbf {\bibinfo {volume} {15}},\ \bibinfo {pages} {055016}
  (\bibinfo {year} {2013})}\BibitemShut {NoStop}%
\bibitem [{Note2()}]{Note2}%
  \BibitemOpen
  \bibinfo {note} {For a realistic spin-degenerate material this one band
  approximation can be realised using the Zeeman splitting associated with a
  magnetic field (since Landau levels will not occur in 1D) or with a
  ferromagnetic substrate.}\BibitemShut {Stop}%
\bibitem [{\citenamefont {Liu}, \citenamefont {Hu},\ and\ \citenamefont
  {Drummond}(2007)}]{Liu-PRA76}%
  \BibitemOpen
  \bibfield  {author} {\bibinfo {author} {\bibfnamefont {X.-J.}\ \bibnamefont
  {Liu}}, \bibinfo {author} {\bibfnamefont {H.}~\bibnamefont {Hu}}, \ and\
  \bibinfo {author} {\bibfnamefont {P.~D.}\ \bibnamefont {Drummond}},\ }\href
  {\doibase 10.1103/physreva.76.043605} {\bibfield  {journal} {\bibinfo
  {journal} {Phys. Rev. A}\ }\textbf {\bibinfo {volume} {76}},\ \bibinfo
  {pages} {043605} (\bibinfo {year} {2007})}\BibitemShut {NoStop}%
\bibitem [{\citenamefont {Parish}\ \emph {et~al.}(2007)\citenamefont {Parish},
  \citenamefont {Baur}, \citenamefont {Mueller},\ and\ \citenamefont
  {Huse}}]{Parish-PRL99}%
  \BibitemOpen
  \bibfield  {author} {\bibinfo {author} {\bibfnamefont {M.~M.}\ \bibnamefont
  {Parish}}, \bibinfo {author} {\bibfnamefont {S.~K.}\ \bibnamefont {Baur}},
  \bibinfo {author} {\bibfnamefont {E.~J.}\ \bibnamefont {Mueller}}, \ and\
  \bibinfo {author} {\bibfnamefont {D.~A.}\ \bibnamefont {Huse}},\ }\href
  {\doibase 10.1103/physrevlett.99.250403} {\bibfield  {journal} {\bibinfo
  {journal} {Phys. Rev. Lett.}\ }\textbf {\bibinfo {volume} {99}},\ \bibinfo
  {pages} {250403} (\bibinfo {year} {2007})}\BibitemShut {NoStop}%
\bibitem [{\citenamefont {Edge}\ and\ \citenamefont
  {Cooper}(2009)}]{Edge-PRL103}%
  \BibitemOpen
  \bibfield  {author} {\bibinfo {author} {\bibfnamefont {J.~M.}\ \bibnamefont
  {Edge}}\ and\ \bibinfo {author} {\bibfnamefont {N.~R.}\ \bibnamefont
  {Cooper}},\ }\href {\doibase 10.1103/physrevlett.103.065301} {\bibfield
  {journal} {\bibinfo  {journal} {Phys. Rev. Lett.}\ }\textbf {\bibinfo
  {volume} {103}},\ \bibinfo {pages} {065301} (\bibinfo {year}
  {2009})}\BibitemShut {NoStop}%
\bibitem [{Note3()}]{Note3}%
  \BibitemOpen
  \bibinfo {note} {This Bessel function diverges as $z\to 0$, so we impose a
  minimum cutoff $|q|d = 10^{-6}$. We have tested that the precise value of
  this cutoff does not impact the numerical results.}\BibitemShut {Stop}%
\bibitem [{\citenamefont {Slachmuylders}\ \emph {et~al.}(2006)\citenamefont
  {Slachmuylders}, \citenamefont {Partoens}, \citenamefont {Magnus},\ and\
  \citenamefont {Peeters}}]{Slachmuylders-PRB74}%
  \BibitemOpen
  \bibfield  {author} {\bibinfo {author} {\bibfnamefont {A.~F.}\ \bibnamefont
  {Slachmuylders}}, \bibinfo {author} {\bibfnamefont {B.}~\bibnamefont
  {Partoens}}, \bibinfo {author} {\bibfnamefont {W.}~\bibnamefont {Magnus}}, \
  and\ \bibinfo {author} {\bibfnamefont {F.~M.}\ \bibnamefont {Peeters}},\
  }\href {\doibase 10.1103/physrevb.74.235321} {\bibfield  {journal} {\bibinfo
  {journal} {Phys. Rev. B}\ }\textbf {\bibinfo {volume} {74}},\ \bibinfo
  {pages} {235321} (\bibinfo {year} {2006})}\BibitemShut {NoStop}%
\bibitem [{\citenamefont {Pistol}\ and\ \citenamefont
  {Pryor}(2008)}]{Pistol-PRB78}%
  \BibitemOpen
  \bibfield  {author} {\bibinfo {author} {\bibfnamefont {M.-E.}\ \bibnamefont
  {Pistol}}\ and\ \bibinfo {author} {\bibfnamefont {C.}~\bibnamefont {Pryor}},\
  }\href {\doibase 10.1103/physrevb.78.115319} {\bibfield  {journal} {\bibinfo
  {journal} {Phys. Rev. B}\ }\textbf {\bibinfo {volume} {78}},\ \bibinfo
  {pages} {115319} (\bibinfo {year} {2008})}\BibitemShut {NoStop}%
\bibitem [{\citenamefont {Kishore}, \citenamefont {Partoens},\ and\
  \citenamefont {Peeters}(2012)}]{Kishore-PRB86}%
  \BibitemOpen
  \bibfield  {author} {\bibinfo {author} {\bibfnamefont {V.~V.~R.}\
  \bibnamefont {Kishore}}, \bibinfo {author} {\bibfnamefont {B.}~\bibnamefont
  {Partoens}}, \ and\ \bibinfo {author} {\bibfnamefont {F.~M.}\ \bibnamefont
  {Peeters}},\ }\href {\doibase 10.1103/physrevb.86.165439} {\bibfield
  {journal} {\bibinfo  {journal} {Phys. Rev. B}\ }\textbf {\bibinfo {volume}
  {86}},\ \bibinfo {pages} {165439} (\bibinfo {year} {2012})}\BibitemShut
  {NoStop}%
\bibitem [{\citenamefont {Efimkin}, \citenamefont {Lozovik},\ and\
  \citenamefont {Sokolik}(2012)}]{Efimkin-PRB86}%
  \BibitemOpen
  \bibfield  {author} {\bibinfo {author} {\bibfnamefont {D.~K.}\ \bibnamefont
  {Efimkin}}, \bibinfo {author} {\bibfnamefont {Y.~E.}\ \bibnamefont
  {Lozovik}}, \ and\ \bibinfo {author} {\bibfnamefont {A.~A.}\ \bibnamefont
  {Sokolik}},\ }\href {\doibase 10.1103/physrevb.86.115436} {\bibfield
  {journal} {\bibinfo  {journal} {Phys. Rev. B}\ }\textbf {\bibinfo {volume}
  {86}},\ \bibinfo {pages} {115436} (\bibinfo {year} {2012})}\BibitemShut
  {NoStop}%
\bibitem [{\citenamefont {Abergel}, \citenamefont {Sensarma},\ and\
  \citenamefont {Das~Sarma}(2012)}]{Abergel-PRB86-excon}%
  \BibitemOpen
  \bibfield  {author} {\bibinfo {author} {\bibfnamefont {D.~S.~L.}\
  \bibnamefont {Abergel}}, \bibinfo {author} {\bibfnamefont {R.}~\bibnamefont
  {Sensarma}}, \ and\ \bibinfo {author} {\bibfnamefont {S.}~\bibnamefont
  {Das~Sarma}},\ }\href {\doibase 10.1103/physrevb.86.161412} {\bibfield
  {journal} {\bibinfo  {journal} {Phys. Rev. B}\ }\textbf {\bibinfo {volume}
  {86}},\ \bibinfo {pages} {161412} (\bibinfo {year} {2012})}\BibitemShut
  {NoStop}%
\bibitem [{\citenamefont {Abergel}\ \emph {et~al.}(2013)\citenamefont
  {Abergel}, \citenamefont {Rodriguez-Vega}, \citenamefont {Rossi},\ and\
  \citenamefont {Das~Sarma}}]{Abergel-PRB88}%
  \BibitemOpen
  \bibfield  {author} {\bibinfo {author} {\bibfnamefont {D.~S.~L.}\
  \bibnamefont {Abergel}}, \bibinfo {author} {\bibfnamefont {M.}~\bibnamefont
  {Rodriguez-Vega}}, \bibinfo {author} {\bibfnamefont {E.}~\bibnamefont
  {Rossi}}, \ and\ \bibinfo {author} {\bibfnamefont {S.}~\bibnamefont
  {Das~Sarma}},\ }\href {\doibase 10.1103/physrevb.88.235402} {\bibfield
  {journal} {\bibinfo  {journal} {Phys. Rev. B}\ }\textbf {\bibinfo {volume}
  {88}},\ \bibinfo {pages} {235402} (\bibinfo {year} {2013})}\BibitemShut
  {NoStop}%
\bibitem [{\citenamefont {{Thelander}}(2014)}]{Thelander}%
  \BibitemOpen
  \bibfield  {author} {\bibinfo {author} {\bibfnamefont {C.}~\bibnamefont
  {{Thelander}}},\ }\href@noop {} {\bibfield  {journal} {\bibinfo  {journal}
  {private communication}\ } (\bibinfo {year} {2014})}\BibitemShut {NoStop}%
\end{thebibliography}%

\end{document}